\documentclass[
amsmath, %
floatfix, %
twocolumn, %
reprint, %
superscriptaddress, %
prb, %
aps, %
citeautoscript, %
final, %
]{revtex4-2} 
\renewcommand{\thispagestyle}[1]{}

\usepackage[T1]{fontenc}
\usepackage[utf8]{inputenc}
\usepackage{graphicx}
\usepackage{latexsym}
\usepackage{amsthm}
\usepackage[low-sup]{subdepth}

\usepackage{color}
\usepackage{mathtools}
\mathtoolsset{showonlyrefs}

\usepackage{placeins}

\usepackage[dvipsnames,table]{xcolor}

\usepackage[]{newtxtext} 
\usepackage[subscriptcorrection,nosymbolsc,smallerops,bigdelims]{newtxmath} 
\DeclareMathAlphabet{\mathcal}{OMS}{cmsy}{m}{n} 
\DeclareMathAlphabet{\mathbcal}{OMS}{cmsy}{b}{n} 
\usepackage{bm}

\usepackage{floatrow}
\usepackage{siunitx}
\usepackage{hyperref}
\hypersetup{
	colorlinks,
	linkcolor={blue!90!black},
	citecolor={blue!90!black},
	urlcolor=	{blue!90!black}
}

\usepackage{multirow}

\let\oldeqref\eqref
\renewcommand*{\eqref}[1]{%
	\hyperref[#1]{\oldeqref{#1}}%
}

\newcommand{\kp}{{\bm{k} {\cdot} \bm{p}}}

\mathchardef\mhyphen="2D

\newcommand{\mr}[1]{\mathrm{#1}}

\DeclarePairedDelimiter\lr{\lparen}{\rparen}

\DeclarePairedDelimiter\abs{\lvert}{\rvert}

\DeclarePairedDelimiterX{\comm}[2]{\lbrack}{\rbrack}{#1, #2}

\DeclarePairedDelimiterX{\braket}[2]{\langle}{\rangle}{#1\delimsize\vert #2}
\DeclarePairedDelimiterX{\ketbra}[2]{\rvert}{\lvert}{#1 \delimsize\rangle\!\delimsize\langle #2}
\DeclarePairedDelimiterX{\matrixel}[3]{\langle}{\rangle}{#1 \delimsize\vert #2 \delimsize\vert #3}

\makeatletter
\newcommand{\raisemath}[1]{\mathpalette{\raisem@th{#1}}}
\newcommand{\raisem@th}[3]{\raisebox{#1}{$#2#3$}}
\makeatother

\definecolor{cbred}{HTML}{e31a1c}
\definecolor{cbgreen}{HTML}{33a02c}
\definecolor{cbblue}{HTML}{176aa7}

\usepackage{dcolumn}
\newcolumntype{d}[1]{D{.}{.}{#1}}

\newcommand{\tabref}[1]{Tab.~\ref{tab:#1}}
\newcommand{\figref}[1]{Fig.~\ref{fig:#1}}
\newcommand{\subfigref}[2]{Fig.~\hyperref[fig:#1]{\ref*{fig:#1}(#2)}}
\newcommand{\subfigsref}[3]{Figs.~\hyperref[fig:#1]{\ref*{fig:#1}(#2)}-\hyperref[fig:#1]{\ref*{fig:#1}(#3)}}

\newcommand{\upcapt}{\vspace{-1.2em}}
\newcommand{\upcaptb}{\vspace{-0.8em}}

\definecolor{cbred}{HTML}{e31a1c}
\definecolor{cbgreen}{HTML}{33a02c}
\definecolor{cbblue}{HTML}{176aa7}
\definecolor{cborange}{HTML}{ff7f00}
\definecolor{cbviolet}{HTML}{6a3d9a}

\usepackage[activate={true,nocompatibility},final,tracking=alltext,kerning=true,spacing=true,protrusion=true,factor=1100,stretch=10,shrink=10,selected=true ,letterspace=-0]{microtype}

\newcommand*{\uPL}{\si{\micro}PL}
\newcommand*{\uPLE}{\si{\micro}PLE}
\newcommand*{\ie}{\textit{i.\,e.}}

\usepackage{array,ragged2e}

\begin{document}
	\title{Excited states of neutral and charged excitons in single strongly asymmetric InP-based nanostructures emitting in the telecom C band}
	
	\author{M.~Gawe{\l}czyk}
	\email{michal.gawelczyk@pwr.edu.pl}
	\affiliation{Department of Theoretical Physics, Faculty of Fundamental Problems of Technology, Wroc\l{}aw University of Science and Technology, 50-370 Wroc\l{}aw, Poland}
	\affiliation{Laboratory for Optical Spectroscopy of Nanostructures, Department of Experimental Physics, Faculty of Fundamental Problems of Technology, Wroc\l{}aw University of Science and Technology, 50-370 Wroc\l{}aw, Poland}
	
	\author{P.~Wyborski}
	\affiliation{Laboratory for Optical Spectroscopy of Nanostructures, Department of Experimental Physics, Faculty of Fundamental Problems of Technology, Wroc\l{}aw University of Science and Technology, 50-370 Wroc\l{}aw, Poland}
	
	\author{P.~Podemski}
	\affiliation{Laboratory for Optical Spectroscopy of Nanostructures, Department of Experimental Physics, Faculty of Fundamental Problems of Technology, Wroc\l{}aw University of Science and Technology, 50-370 Wroc\l{}aw, Poland}	
	
	\author{J.\,P.~Reithmaier}
	\affiliation{Institute of Nanostructure Technologies and Analytics (INA), CINSaT, University of Kassel, Heinrich-Plett-Str.\ 40, 34132 Kassel, Germany}
	
	\author{S.~H\"{o}fling}
	\affiliation{Technische Physik, University of W\"{u}rzburg and Wilhelm-Conrad-R\"{o}ntgen-Research Center for Complex Material Systems, Am Hubland, D-97074 W\"{u}rzburg, Germany}
	\affiliation{SUPA, School of Physics and Astronomy, University of St.Andrews, North Haugh, KY16 9SS St.\ Andrews, United Kingdom}
	
	\author{G.~S{\k{e}}k}
	\affiliation{Laboratory for Optical Spectroscopy of Nanostructures, Department of Experimental Physics, Faculty of Fundamental Problems of Technology, Wroc\l{}aw University of Science and Technology, 50-370 Wroc\l{}aw, Poland}
	
	\begin{abstract}
		We investigate strongly asymmetric self-assembled nanostructures with one of dimensions reaching hundreds of nanometers.
		Close to the nanowire-like type of confinement, such objects are sometimes assigned as one-dimensional in nature.
		Here, we directly observe the spectrum of exciton excited states corresponding to longitudinal quantization.
		This is based on probing the optical transitions via polarization-resolved microphotoluminescence excitation (\si{\micro}PLE) measurement performed on single nanostructures combined with theoretical calculation of neutral and charged exciton optical properties.
		We successfully probe absorption-like spectra for individual bright states forming the exciton ground-state fine structure, as well as for the negatively charged exciton.
		Confronting the calculated spectrum of excitonic absorption with \si{\micro}PLE traces, we identify optical transitions involving states that contain carriers at various excited levels related to the longest dimension.
		Based on cross-polarized excitation-detection scheme, we show very well conserved spin configuration during orbital relaxation of the exciton from a number of excited states comparable to the quasi-resonant pumping via the optical phonon, and no polarization memory for the trion, as theoretically expected.
	\end{abstract}
	
	\maketitle

	Out of the abundance of semiconductor nanostructures, those prepared via molecular beam epitaxy (MBE) in InAs on InP-substrate systems \cite{BraultAPL1998}, especially InAs/AlGaInAs considered here \cite{SauerwaldAPL2005}, typically stand out with high areal density and in-plane elongation so high that it is disputable whether they should be classified as quantum dots (QDs) or quantum wires \cite{DeryJAP2004,KhanPiQE2014,RyczkoJoAP2014,GawelczykPRB2017}.
	This fundamental ambiguity did not prevent the intense research on the properties of such objects.
	They have been studied both as ensembles \cite{DeryJAP2004,ReithmaierJPhysD2008,CapuaOptExpress2012,SyperekAPL2013} and single objects \cite{SekJAP2009,DusanowskiAPL2013,DusanowskiAPL2016,DusanowskiAPL2018} showing potential for implementation in optoelectronics like telecom lasers and optical amplifiers \cite{HeinEL2008,CapuaAPL2010,CapuaOptExpress2012,CapuaPRB2014} or single-photon sources for fiber networks \cite{DusanowskiAPL2014,DusanowskiAPL2016}.
	Notably, their emission energy is tuned deterministically over the telecom C band with the amount of deposited InAs \cite{SauerwaldAPL2005}.
	
	Several other strategies for achieving telecom emission from epitaxial QDs are actively developed including sophisticated approaches like deposition on metamorphic buffers or droplet epitaxy. These proved successful on both GaAs and InP substrates \cite{LiuPRB2014,HaJJoAP2015,HaAPE2016,SkibaSzymanskaPRA2017} and yielded low or vanishing exciton fine structure splitting (FSS). In parallel, efforts aiming at improving properties of self-assembled MBE-grown InP-based nanostructures like those considered here continue. 
	Cancellation of FSS is achievable also in this case via the lately proposed approach \cite{MrowinskiAPL2015}.
	
	Regarding basic research recent studies brought understanding of exciton ground-state (GS) properties in investigated QDs including partially polarized emission \cite{MusialPRB2012}, two-exponential recombination \cite{GawelczykPRB2017}, and raised questions about exciton confinement regime \cite{DusanowskiAPL2017,GawelczykAPPA2018a}.
	Although excited exciton states in QDs were the subjectc of a recent deep study \cite{HoltkemperPRB2018}, the realm of highly elongated dots remains weakly covered in this context, with a single work for InAs/InP QDs \cite{ZielinskiPRB2019}, where, however, the confinement is different.
	The available QD-ensemble absorption data \cite{MarynskiJoAP2013,RudnoRudzinskiAPL2006} is insufficient, as energy separations below the ensemble bandwidth are expected for quantization in the longest dimension \cite{GawelczykPRB2017}.
	Thus, high-resolution photoluminescence excitation (PLE) spectroscopy measurements are needed.
	These are widely used for QDs emitting below $\SI{1}{\micro\meter}$, but are experimentally challenging in the telecom range.
	One remedy is to use pulsed excitation \cite{DusanowskiOE2017}.
	This however provides low average power and can increase linewidth, which obstructs resolution of individual transitions, ladder of which is dense for investigated QDs.
	
	Here, we use an experimental setup built for PLE studies of single nanostructures (\uPLE) emitting in the infrared \cite{PodemskiJoL2019}, especially in the third telecom window.
	We successfully probe the spectrum of optically active excited states of C-band-emitting QDs elongated above $\SI{100}{\nano\metre}$, including individual bright states forming the exciton fine structure, and the negatively charged exciton (X$^{-}$).
	Confrontation of this \uPLE{} data with calculated absorption spectra indicates the presence of states that involve electron and hole excitations related to the longest QD dimension.
	With this, we unambiguously confirm the zero-dimensional nature of studied objects, \ie, energy quantization in all spatial dimensions.
	While additional localization in quantum wires is known to lead to energy quantization \cite{IntontiPRB2001}, here we demonstrate quantization in the longest QD dimension exceeding 100 nm although no further effective size reduction occurs.
	The agreement of simulated and experimental spectra allows us to label the transitions with electron-hole orbital-state configurations forming respective excited states.
	These are dominated by highly-excited hole state contributions due to weak hole confinement in the system.
	Additionally, in a polarization-resolved excitation-detection scheme, we observe mostly conserved spin configuration during orbital relaxation from a number of states for the neutral exciton and, contrarily, no spin-memory for X$^{-}$ in agreement with theoretical considerations.
	
	The investigated sample was grown by gas-source MBE on InP:S substrate \cite{SauerwaldAPL2005}.
	Deposition of \SI{1.26}{\nano\meter} of InAs onto a \SI{200}{\nano\meter} thick Al$_{0.24}$Ga$_{0.23}$In$_{0.53}$As barrier layer yielded a Stranski-Krastanov formation of nanostructures on a wetting layer (WL), covered with \SI{100}{\nano\meter} thick barrier layer and \SI{20}{\nano\meter} of InP.
	Imaging of the uncapped sample showed a dense ($\mathbin{>}\SI{e10}{\centi\meter^{-2}}$) ensemble of nanostructures strongly elongated in the $v\mathbin{\equiv}[1\bar{1}0]$ direction ($h\mathbin{\equiv}[110]$).
	Cross sections are triangle-like with a width-to-height ratio of $W/H\approx6$ \cite{SauerwaldAPL2005}.
	The length $L$ may significantly exceed \SI{100}{\nano\metre} \cite{RudnoRudzinskiAPL2006,MusialPRB2012}, which places the structures on a crossover between dot- and wire-like confinements.
	To enable optical experiments on single nanostructures, the sample was processed by electron beam lithography and etching, which left sub-micrometer mesas.
	
	During experiments the sample was kept in a continuous-flow liquid-helium cryostat at $T\mathbin{=}\SI{4.2}{\kelvin}$.
	Structures were excited either nonresonantly with a continuous-wave (CW) \SI{639}{\nano\metre} laser line, or quasi-resonantly with an external-cavity CW laser in the Littrow configuration with a tuning range of $\SIrange{1440}{1540}{\nano\metre}$.
	The excitation beam was filtered with a shortpass filter and a 300-mm-focal-length monochromator to improve its quality.
	Spatial resolution of $\SI{\sim2}{\micro\metre}$ was obtained by passing both the excitation beam and the collected signal through a microscope objective with 0.4 numerical aperture.
	After removing the scattered laser line with a longpass or bandpass filter, the collected signal was guided to an analyzer consisting of a 1-m-focal-length monochromator and a GaInAs-based multichannel detector with a spectral resolution of $\SI{\sim50}{\micro\electronvolt}$.
	Rotation of a half-wave plate in front of a linear polarizer allowed for polarization resolution in both excitation and collection.
	
	\begin{figure}[!t] %
	\begin{center} %
		\includegraphics[width=\columnwidth]{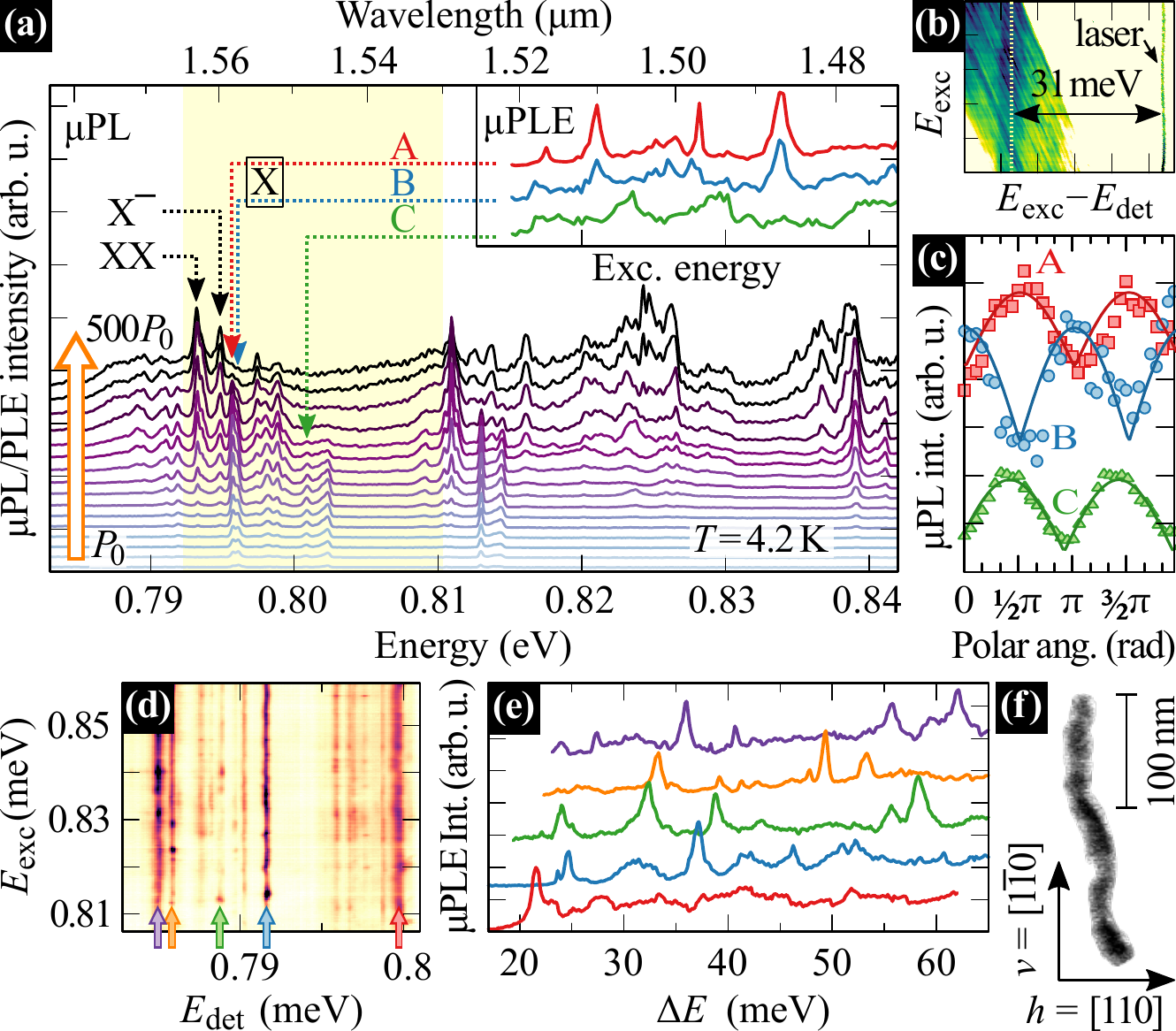} %
	\end{center} %
	\upcapt\caption{\label{fig:PL-PLE}(a) Power series of \uPL\ spectra.
		Inset: \uPLE\ collected at energies marked with arrows, energy axis is common.
		(b) PLE map from ensemble of QDs; LO-phonon line marked.
		(c) Polarization-resolved \uPL\ with $a\,\abs{\sin{x}}$ fits for lines A--C.
		(d) \uPLE\ color map for an exemplary mesa with QDs.
		(e) \uPLE\ spectra collected at energies marked on the map.
		(f) Single QD top view extracted from scanning electron microscopy image of an uncapped sample.\upcaptb} %
	\end{figure}	

	In \subfigref{PL-PLE}{a}, we plot microphotoluminescence (\uPL) spectra collected from a selected $550\times 275\:\si{\nano\meter\squared} \approx \SI{0.15}{\micro\meter\squared}$ mesa at various power densities $P$ of nonresonant CW excitation ($P_0\mathbin{\sim}\SI{30}{\nano\watt\per\micro\meter^2}$ in the spot).
	Those from observed well-resolved lines covering the C band that occur already at the lowest power come from carrier-complex GSs.
	We choose three marked with arrows for further analysis, based on their linear growth.
	The polarization-resolved \uPL\ shows no fine structure splitting within the available resolution for line C, which thus comes most probably from X$^{-}$, as the positively-charged exciton is less probable to be formed in this system \cite{MrowinskiPRB2016,ZielinskiPRB2019}. Studied QDs, due to their asymmetry, show FSS in the range of $\SIrange{50}{180}{\micro\electronvolt}$ \cite{MrowinskiPRB2016}, so our spectral resolution along with polarization properties of the line allow for this assignment.
	Contrarily, lines A and B come from two spin configurations of the same exciton GS with $\mr{FSS}\mathbin{\simeq}\SI{240}{\micro\electronvolt}$, which exceeds previous observations \cite{MrowinskiPRB2016,ZielinskiPRB2019}. In \subfigref{PL-PLE}{b}, we show intensities of these lines versus polarization angle relative to $[110]$, where A and B are in antiphase, as expected. See Supplemental Material [\onlinecite{Supplemental}] for a more detailed analysis of the A-B-XX recombination cascade.\vphantom{\cite{SekJoAP2010,GrundmannPRB1997,AbbarchiJoAP2009}}
	
	Additionally, in the inset to \subfigref{PL-PLE}{a}, \uPLE\ spectra collected at energies corresponding to chosen emission lines are shown.
	Noticeably, traces for A and B are similar, which confirms their classification.
	To support the analysis, we show in \subfigref{PL-PLE}{b} a QD-ensemble PLE map, with a line occurring at fixed $\SI{\sim31}{\milli\electronvolt}$ separation from the laser, identified as the LO-phonon-assisted absorption \cite{LemaitrePRB2001}.
	Apparently, the volume of QDs is large enough to modify phonon dispersion, as the LO phonon energy in AlGaInAs is $E_{\mr{LO}}\SI{\sim36}{\milli\electronvolt}$, and the one observed is closer to the InAs value.
	
	Next, in \subfigref{PL-PLE}{d}, we present a typical excitation-detection map, where each vertical cross-section forms a \uPLE\ spectrum.
	Arrows mark lines, spectra for which are plotted in \subfigref{PL-PLE}{e}.
	Their significant diversity reflects the inhomogeneity of QDs in question: spread lengths, and factors like local widenings and bends visible in \subfigref{PL-PLE}{f}.
	Note that the height of PLE peaks results from both absorption rate and effective orbital relaxation rate for given $\Delta E$.
	
\begin{figure}[!t] %
	\begin{center} %
		\includegraphics[width=\columnwidth]{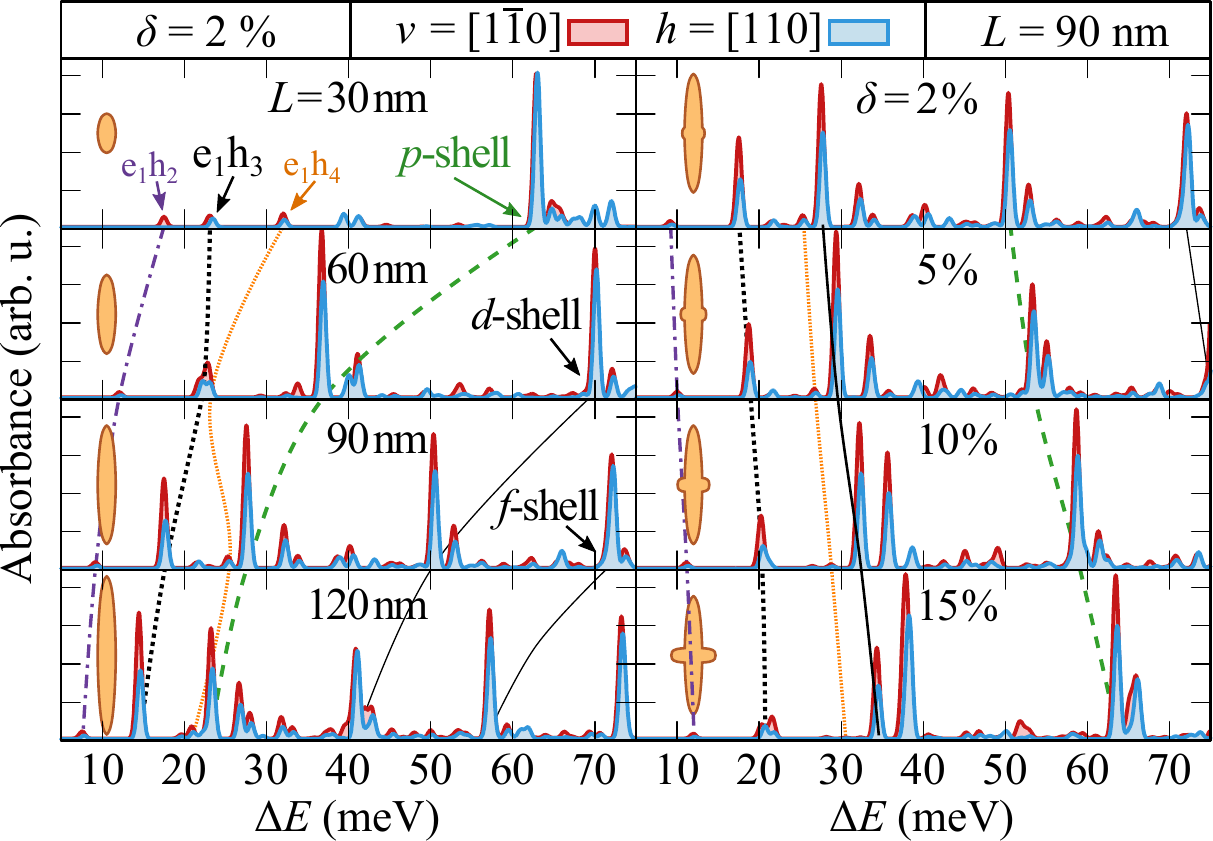} %
	\end{center} %
	\upcapt\caption{\label{fig:absorption-L-eta}Calculated polarization-resolved neutral-exciton absorption spectra for various QD lengths $L$ (left) and hump sizes $\delta$ (right). Insets: schematic QD geometry.\upcaptb} %
	\end{figure}

	To understand the \uPLE\ data, we calculate optical properties of excited states for QDs of various geometry.
	Protruding from a \SI{0.9}{\nano\metre} thick WL, modeled QDs have triangular cross section, $W/H\mathbin{=}6$, and elliptical longitudinal height profile.
	To account for material intermixing, we perform Gaussian averaging with $\sigma\mathbin{=}\SI{0.9}{\nano\metre}$ \cite{GawelczykPRB2017}.
	Apart from varying $H$ in the range of $\SIrange{1.8}{3.3}{\nano\metre}$ and $L$ within $\SIrange{30}{280}{\nano\metre}$, we account for expected perturbed geometry (see \subfigref{PL-PLE}{e} and Ref.~[\onlinecite{MusialPRB2012}]) via a \SI{10}{\nano\metre} long central hump with $\lr{1+\delta}$ times enlarged cross-section (see insets in \figref{absorption-L-eta}).
	Conduction- and valence-band electron states are calculated using the 8-band envelope-function $\kp$ theory \cite{BahderPRB1990,BurtJPhysCondMat1992,ForemanPRB1993}.
	We use a numerical implementation \cite{GawareckiPRB2014} that includes spin-orbit effects, strain and nonlinear piezoelectric field (for details and parameters see Ref.~[\onlinecite{GawelczykPRB2017}]).
	Then, we construct complex states via diagonalization of Coulomb and anisotropic electron-hole exchange interactions within the configuration-interaction approach ($46\times46$ single-particle configurations used) and calculate polarization-resolved oscillator strengths in the dipole approximation \cite{AndrzejewskiJAP2010}.
	To simulate absorption spectra, we widen each line by $\Delta E \mathbin{=} \hbar / \tau$ ($\tau$ is the calculated lifetime), and convolute with a Gaussian (and its LO-phonon replicas) of $\sigma\mathbin{=}\SI{0.2}{\milli\electronvolt}$ to represent the laser linewidth, spectral diffusion and phonon effects \cite{DusanowskiPRB2014}.
	Where compared with experiment, curves are scaled with $\cramped{1-\exp\lr{-ax}}$ to account for state saturation.

	Calculated absorption spectra are plotted in \figref{absorption-L-eta}.
	While $H$ determines GS energy $E_0$, it weakly affects relative positions of states, thus we present only results for $H\mathbin{=}\SI{3.3}{\nano\metre}$ corresponding to $E_{\mr{X}}\mathbin{\sim}\SI{0.8}{\electronvolt}$: for QDs with $\delta\mathbin{=}\SI{2}{\percent}$ and varying $L\mathbin{=}\SIrange{30}{120}{\nano\metre}$ (left), and for fixed $L\mathbin{=}\SI{90}{\nano\metre}$ with varied $\delta$ (right).
	The two curves represent absorption of $v/h$-polarized light.
	Phonon replicas are disabled here for clarity.
	Understandably, with rising $L$ the spectrum gets denser.
	The hump reduces $E_0$, thus shifts excited states to higher relative energies without much change in their spacing.
	The dashed line traces the $p$-shell redshift following a $\mathbin{\sim}1/L$ trend that has been already noticed \cite{GawelczykPRB2017}.
	Additionally, another significantly bright state emerges with QD length below $p$-shell.
	It is predominantly composed of the GS electron and the hole at third orbital level (three antinodes along the QD).
	The high confinement anisotropy results in nearly independent sub-ladders of excitations in in-plane directions with level spacings defined by $L$ and $H$ \cite{GawelczykPRB2017}. 
	Effectively heavier holes experience shallow confinement in this material system, hence spacing of their longitudinal excitations is on the level of single meVs.
	This underlies the presence of the bright state below $p$-shell.
	We additionally trace higher shells and two nominally dark states $\mr{e}_1\mr{h}_2,~\mr{e}_1\mr{h}_4$.
	All states have two bright spin configurations coupling to $v$- and $h$-polarized light.
	
	\begin{figure}[!t] %
	\begin{center} %
		\includegraphics[width=\columnwidth]{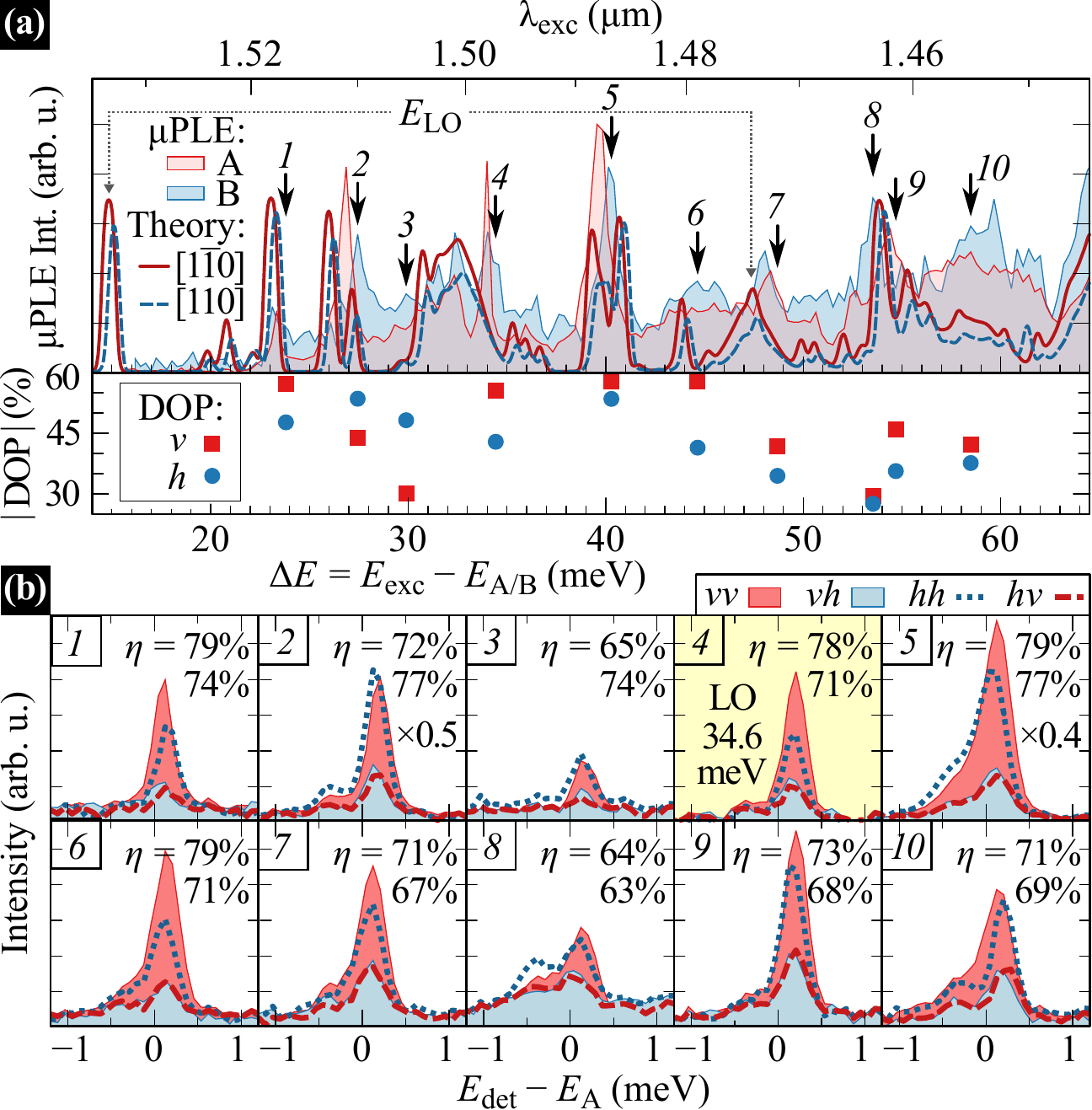} %
	\end{center} %
	\upcapt\caption{\label{fig:PLE-polar-X}(a) \uPLE\ spectra for the exciton fine structure pair of lines [A and B in \subfigref{PL-PLE}{a}] (filled) and calculated polarization-resolved absorption spectra for QD with $L\mathbin{=}\SI{140}{\nano\metre},~\delta\mathbin{=}\SI{5}{\percent}$ (lines); absolute values of DOP for emission under $v$- and $h$-polarized excitation at marked peaks (symbols, bottom panel). (b) \uPL\ of given lines for excitation at marked \uPLE\ maxima in various configurations of excitation-detection polarization.\upcaptb} %
	\end{figure}	
	
	Returning to lines A and B from \subfigref{PL-PLE}{a}, in \subfigref{PLE-polar-X}{a} we confront their \uPLE\ traces with calculated absorption spectra.
	Judging by the agreement, the line likely comes from a QD well modeled by $L\mathbin{=}\SI{140}{\nano\metre}$ and $\delta\mathbin{=}\SI{5}{\percent}$.
	Here, we set $E_{\mr{LO}}\mathbin{=}\SI{32.5}{\milli\electronvolt}$, based on visible wide phonon replicas.
	The one sharp line at $\SI{\sim34}{\milli\electronvolt}$ without a counterpart in theory is likely a bulk phonon replica, as opposed to wide peaks originating from perturbed phonon modes.
	While the calculated position of $\mr{e}_1\mr{h}_3$ state is below the range of experimental data, its replica fits well the widened experimental peak at $\Delta E\mathbin{\simeq}\SI{48}{\milli\electronvolt}$, which, according to calculation, is formed by overlapping $\mr{e}_1\mr{h}_3$ replica and another transition. 
	We characterize also other excited states corresponding to marked peaks, as presented in \tabref{states-composition} via a few dominant single-particle contributions.
	We label them by the state number $n$ and axis-wise excitations $n_{v/h}$, \ie, numbers of antinodes along $v$ and $h$.
	Noticeably, no purely $p$-shell state is present, as it got mixed with $\mr{e}_1\mr{h}_3$ (first column).
	While all states are predominantly composed of the GS electron configurations with weaker admixtures of few excited levels, the hole contributes significantly with a number of excited states: both sub-ladders of states are involved, with the dominance of longitudinal excitations. 
	Analogous discrete absorption spectrum could not be obtained assuming the quantum-wire confinement limit. For more examples of theoretically reproduced \uPLE\ spectra, see Supplemental Material [\onlinecite{Supplemental}].

	\begin{ruledtabular}
	\begingroup \squeezetable
	\begin{table}[!t]
		\sisetup{round-mode=figures,round-precision=2}
		\setlength\tabcolsep{0.pt}
		\begin{tabular*}{\textwidth}{@{\extracolsep{\fill}}c<{\,} | rrrrrrrrr}\rule{0pt}{1.15em}
			$\Delta E_{\mr{exp}}$\,(meV)	&	23.8				\cellcolor{cbblue!	0	} &	27.4				\cellcolor{cbblue!	0	} &	29.9				\cellcolor{cbblue!	0	} &	40.3				\cellcolor{cbblue!	0	} &	44.7				\cellcolor{cbblue!	0	} &	48.9				\cellcolor{cbblue!	0	} &	53.5				\cellcolor{cbblue!	0	} &	54.7				\cellcolor{cbblue!	0	} &	58.5				\cellcolor{cbblue!	0	}\\
			$\Delta E_{\mr{calc}}$\,(meV)	&	23.1				\cellcolor{cbblue!	0	} &	27.2				\cellcolor{cbblue!	0	} &	29.5				\cellcolor{cbblue!	0	} &	40.6				\cellcolor{cbblue!	0	} &	43.9				\cellcolor{cbblue!	0	} &	47.4				\cellcolor{cbblue!	0	} &	53.9				\cellcolor{cbblue!	0	} &	55.1				\cellcolor{cbblue!	0	} &	58.1				\cellcolor{cbblue!	0	}\\
			$f_v$	&	\num{6.504}				\cellcolor{cbblue!	0	} &	\num{1.082}				\cellcolor{cbblue!	0	} &	\num{0.083}				\cellcolor{cbblue!	0	} &	\num{3.798}				\cellcolor{cbblue!	0	} &	\num{0.667}				\cellcolor{cbblue!	0	} &	\num{0.249}				\cellcolor{cbblue!	0	} &	\num{2.944}				\cellcolor{cbblue!	0	} &	\num{0.587}				\cellcolor{cbblue!	0	} &	\num{0.209}				\cellcolor{cbblue!	0	}\\
			$f_h$	&	\num{4.040}				\cellcolor{cbblue!	0	} &	\num{0.698}				\cellcolor{cbblue!	0	} &	\num{0.064}				\cellcolor{cbblue!	0	} &	\num{3.336}				\cellcolor{cbblue!	0	} &	\num{0.386}				\cellcolor{cbblue!	0	} &	\num{0.202}				\cellcolor{cbblue!	0	} &	\num{2.570}				\cellcolor{cbblue!	0	} &	\num{0.345}				\cellcolor{cbblue!	0	} &	\num{0.196}				\cellcolor{cbblue!	0	}\\\hline
			&	1:\,1|1	\cellcolor{cbblue!	22.5	} &	1:\,1|1	\cellcolor{cbblue!	24	} &	1:\,1|1	\cellcolor{cbblue!	41.5	} &	1:\,1|1	\cellcolor{cbblue!	19.35	} &	1:\,1|1	\cellcolor{cbblue!	21.5	} &	1:\,1|1	\cellcolor{cbblue!	21.325	} &	1:\,1|1	\cellcolor{cbblue!	19.4	} &	1:\,1|1	\cellcolor{cbblue!	15.35	} &	1:\,1|1	\cellcolor{cbblue!	16.9	}\\
			&	2:\,2|1	\cellcolor{cbblue!	22.5	} &	2:\,2|1	\cellcolor{cbblue!	17.5	} &	2:\,2|1	\cellcolor{cbblue!	6.8	} &	2:\,2|1	\cellcolor{cbblue!	15.1	} &	2:\,2|1	\cellcolor{cbblue!	19.95	} &	2:\,2|1	\cellcolor{cbblue!	13.95	} &	3:\,3|1	\cellcolor{cbblue!	11.2	} &	3:\,3|1	\cellcolor{cbblue!	14.025	} &	3:\,3|1	\cellcolor{cbblue!	11.25	}\\
			&	3:\,3|1	\cellcolor{cbblue!	4	} &	3:\,3|1	\cellcolor{cbblue!	6.2	} &					\cellcolor{cbblue!	0	} &	3:\,3|1	\cellcolor{cbblue!	8.95	} &	3:\,3|1	\cellcolor{cbblue!	5.75	} &	3:\,3|1	\cellcolor{cbblue!	7.75	} &	4:\,4|1	\cellcolor{cbblue!	10.8	} &	2:\,2|1	\cellcolor{cbblue!	13.325	} &	4:\,4|1	\cellcolor{cbblue!	9.6	}\\
			\multirow{-4}{*}{\shortstack{e states\\contribution\\$n{:}\,n_v\vert\, n_h$}}	&					\cellcolor{cbblue!	0	} &					\cellcolor{cbblue!	0	} &					\cellcolor{cbblue!	0	} &	4:\,4|1	\cellcolor{cbblue!	4.6	} &					\cellcolor{cbblue!	0	} &	4:\,4|1	\cellcolor{cbblue!	6.95	} &	2:\,2|1	\cellcolor{cbblue!	4.05	} &	4:\,4|1	\cellcolor{cbblue!	4.25	} &	2:\,2|1	\cellcolor{cbblue!	8.2	}\\\hline
			&	2:\,2|1	\cellcolor{cbred!	19.5	} &	4:\,4|1	\cellcolor{cbred!	17.8	} &	9:\,2|2	\cellcolor{cbred!	34.75	} &	7:\,6|1	\cellcolor{cbred!	9.35	} &	12:\,9|1	\cellcolor{cbred!	13	} &	19:\,2|3	\cellcolor{cbred!	9.525	} &	4:\,4|1	\cellcolor{cbred!	10.9	} &	8:\,7|1	\cellcolor{cbred!	11	} &	23:\,13|1	\cellcolor{cbred!	15.475	}\\
			&	3:\,3|1	\cellcolor{cbred!	18.7	} &	3:\,3|1	\cellcolor{cbred!	13	} &	6:\,1|2	\cellcolor{cbred!	6.4	} &	12:\,9|1	\cellcolor{cbred!	6.5	} &	5:\,5|1	\cellcolor{cbred!	8.5	} &	5:\,5|1	\cellcolor{cbred!	5.85	} &	20:\,12|1	\cellcolor{cbred!	11.1	} &	20:\,12|1	\cellcolor{cbred!	8.05	} &	5:\,5|1	\cellcolor{cbred!	7.45	}\\
			&	1:\,1|1	\cellcolor{cbred!	6.4	} &	2:\,2|1	\cellcolor{cbred!	6.8	} &	13:\,4|2	\cellcolor{cbred!	3.2	} &	3:\,3|1	\cellcolor{cbred!	6.2	} &	9:\,2|2	\cellcolor{cbred!	6.25	} &	7:\,6|1	\cellcolor{cbred!	5.4	} &	22:\,3|3	\cellcolor{cbred!	5.75	} &	12:\,9|1	\cellcolor{cbred!	6.05	} &	7:\,6|1	\cellcolor{cbred!	6.225	}\\
			&					\cellcolor{cbred!	0	} &	1:\,1|1	\cellcolor{cbred!	3	} &					\cellcolor{cbred!	0	} &	4:\,4|1	\cellcolor{cbred!	5.3	} &	7:\,6|1	\cellcolor{cbred!	5.75	} &	2:\,2|1	\cellcolor{cbred!	4.7	} &	5:\,5|1	\cellcolor{cbred!	5.35	} &	10:\,8|1	\cellcolor{cbred!	5.7	} &	8:\,7|1	\cellcolor{cbred!	4.025	}\\
			\multirow{-5}{*}{\shortstack{h states\\contribution\\$n{:}\,n_v\vert\, n_h$}}	&					\cellcolor{cbred!	0	} &					\cellcolor{cbred!	0	} &					\cellcolor{cbred!	0	} &	5:\,5|1	\cellcolor{cbred!	3.8	} &	10:\,8|1	\cellcolor{cbred!	2.5	} &	14:\,10|1	\cellcolor{cbred!	2.95	} &	7:\,6|1	\cellcolor{cbred!	4.15	} &	7:\,6|1	\cellcolor{cbred!	4	} &	12:\,9|1	\cellcolor{cbred!	3.25	}
		\end{tabular*} \upcaptb\upcaptb
		\caption{\label{tab:states-composition}Calculated energy, oscillator strengths $f_{v/h}$, and single-particle components of exciton states (contribution size indicated by color intensity) matching selected transitions in \figref{PLE-polar-X}.}
	\end{table}\endgroup
	\end{ruledtabular}

\begin{figure}[!t] %
	\begin{center} %
		\includegraphics[width=\columnwidth]{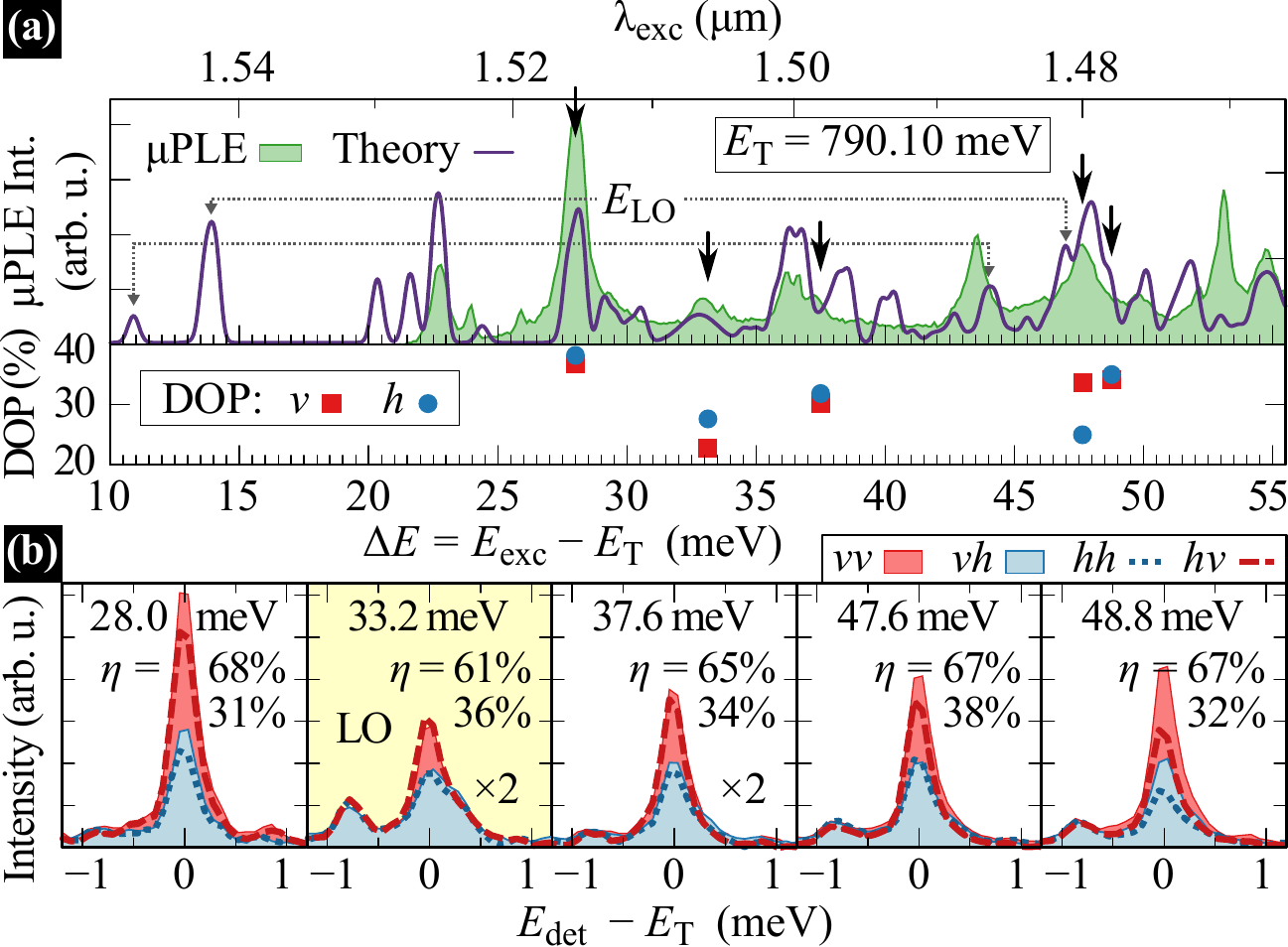} %
	\end{center} %
	\upcapt\caption{\label{fig:PLE-polar-T}As in \figref{PLE-polar-X} but for X$^{-},~L\mathbin{=}\SI{140}{\nano\metre},~\delta\mathbin{=}\SI{10}{\percent}$.} %
	\end{figure}

	Inefficient spin relaxation in QDs should allow for preservation of linear polarization between pumping into excited exciton state and GS emission.
	To verify this, we present in \subfigref{PLE-polar-X}{b} \uPL\ from lines A and B in four configurations of excitation-detection linear polarization and under pumping into selected \uPLE\ peaks.
	The highlighted panel corresponds to the peak identified as an overlap of LO-phonon-assisted GS- and excited-state absorption.
	Based on expected high efficiency of the former, which should be spin-preserving, we use the corresponding values of polarization-injection efficiency $\eta_{v/h}\mathbin{=}I_{v/h}/(I_{v/h}+I_{h/v})$ as a reference.
	These, $\eta_v^{(\mr{LO})}\mathbin{=}\SI{78}{\percent},~\eta_h^{(\mr{LO})}\mathbin{=}\SI{71}{\percent}$, do not stand out from those obtained for other lines.
	Thus, exciton relaxation through the ladder of states in the probed range of energies is highly spin-preserving.
	While higher than those for quasi-resonant pumping of QD ensemble \cite{SyperekAPL2016}, the values of $\eta$ are subideal, partly due to unavoidable misalignment of misshapen QDs with respect to polarization axes.
	In \subfigref{PLE-polar-X}{a} we replot this data in the form of $v$-axis degree of polarization, $\mr{DOP}=\lr{I_v-I_h}/\lr{I_v+I_h}$, of emission under excitation with each of polarizations (for $h$-pumping $\mr{DOP}<0$, and we show $\abs{\mr{DOP}}$),
	
	Next, we repeat the analysis as above for the X$^{-}$ line from another mesa.
	In \subfigref{PLE-polar-T}{b}, we plot the corresponding \uPLE\ trace with agreeing calculated absorption spectrum, achieved for $L\mathbin{=}\SI{140}{\nano\metre},~\delta\mathbin{=}\SI{10}{\percent}$.
	Notably, the low-energy states predicted by calculation below the experimental range have phonon replicas in the measured \uPLE\ signal.
	Apart from denser state ladder due to three-particle nature of the complex, the main difference relative to the exciton is the absence of fine structure and resultant linear polarization of absorbed light.
	Here, bright spin configurations couple to light polarized elliptically, with major axes inclined towards $v$ for both states.
	Consequently, the two states should get equally occupied under both pumping polarizations, and emit with elliptical one, with linear projections equally unequal for both states.
	This is revealed in polarization-resolved \uPL\ from the given line under polarized excitation presented in \subfigref{PLE-polar-T}{b}, where approximately no impact of pumping beam polarization on the emission is observed, and reflected in DOP plotted in \subfigref{PLE-polar-T}{a}.
	Values of $\eta_v$ are misleading here, as the states naturally emit partly polarized light.
	Thus, as the elliptical basis is nonorthogonal, polarized light cannot be used to selectively address X$^{-}$ states.\\[1em]

	In conclusion, we have studied the spectrum of optically active excited states in QDs of InP-based material system characterized by strong in-plane anisotropy and emitting in the telecom C band.
	We have used an experimental setup with filtered tunable external-cavity laser excitation providing high spatial resolution and tuned excitation in the infrared.
	Performing \uPLE\ experiments on single QDs, we have obtained absorption-like spectra of neutral and charged excitons.
	Combining this with calculations, and based on the agreement, we have identified transitions involving states that contain carriers at various excitation levels related to the longest QD dimension.
	This confirms the zero-dimensional character of carrier confinement in studied QDs, which are often treated and modeled as one-dimensional quantum wires.
	Additionally, using a cross-polarized excitation-detection scheme, we have shown highly spin-preserving exciton relaxation in a range of excitation energies.
	Contrarily, but in line with theory, the charged exciton showed no such linear-polarization memory.
	
	\begin{acknowledgments}
		We acknowledge support from the National Science Centre (Poland) under Grants Nos. 2014/14/M/ST3/00821 and 2014/15/D/ST3/00813.
		The project has also been financially supported by the Polish National Agency for Academic Exchange. P.\,W. acknowledges support by the European Union under the European Social Fund.
		S.\,H. acknowledges the support from the State of Bavaria in Germany.
		Numerical calculations have been carried out using resources provided by Wroclaw Centre for Networking and Supercomputing \cite{wcss}, Grant No.~203.
		We are grateful to Krzysztof Gawarecki for sharing his implementation of the $\kp$ method, and Pawe\l\ Machnikowski for helpful discussions.
	\end{acknowledgments}
	\FloatBarrier

%

\end{document}